\documentstyle[12pt]{elsart}
\begin{document}
\newcommand{\eq}{\begin{eqnarray}}
\newcommand{\en}{\end{eqnarray}}
\newcommand{\co}{\, ,} 
\newcommand{\per}{\, .}
\begin{frontmatter}

\begin{flushright}
{BUTP-99/20}
\end{flushright}

\title{              
Numerical analysis of the $\pi^+\pi^-$ atom lifetime in ChPT}

\author[Bern]{J. Gasser}, 
\author[Bern,Dubna,Tomsk]{V.E. Lyubovitskij} and
\author[Bern,Dubna,Tbilisi]{A. Rusetsky}
\address[Bern]{Institute for Theoretical Physics, University of Bern,
Sidlerstrasse 5, CH-3012, Bern, Switzerland}
\address[Dubna]{Bogoliubov Laboratory of Theoretical Physics, Joint Institute
for Nuclear Research, 141980 Dubna, Russia}
\address[Tomsk]{Department of Physics, Tomsk State University, 
634050 Tomsk, Russia}
\address[Tbilisi]{HEPI, Tbilisi State University, 380086 Tbilisi, Georgia}

\date{21 October 1999}

\maketitle

{\footnotesize{e-mails: gasser@itp.unibe.ch, \hspace{2mm} lubovit@itp.unibe.ch,
    \hspace{2mm} rusetsky@itp.unibe.ch}}

\vskip.5cm

\begin{abstract}
We apply Chiral Perturbation Theory at one loop to analyze the general 
formula for the $\pi^+\pi^-$ atom lifetime derived recently in the 
framework of QCD~\cite{Bern}. The corresponding analytic expression is 
investigated numerically, and compared with recent work in the literature.

\vskip1cm


\vskip.5cm


\end{abstract}

\end{frontmatter}

{\bf 1.} The DIRAC collaboration at CERN~\cite{DIRAC} aims to measure the 
lifetime of the $\pi^+\pi^-$ atom (pionium) in its ground state at the 
10\% level. This atom decays predominantly into two neutral pions,
$\Gamma=\Gamma_{2\pi^0}+\Gamma_{2\gamma}+\ldots,\;$ with 
$\Gamma_{2\gamma}/\Gamma_{2\pi^0}\sim 4\cdot 10^{-3}$~\cite{DIRAC}.
The measurement of $\Gamma_{2\pi^0}$ allows 
one~\cite{Bern,Deser,Uretsky,Bilenky} to determine the difference $a_0-a_2$ 
of the strong $S$-wave $\pi\pi$ scattering lengths with isospin $I=0,2$. 
One may then confront the  predictions for this quantity obtained in 
standard ChPT~\cite{ChPTlit,ChPT} with the lifetime measurement, and  
furthermore analyze the nature of spontaneous chiral symmetry breaking in 
QCD~\cite{Stern}. In order to perform these investigations, one needs to 
know the theoretical expression for the width of pionium with a precision 
that properly matches the accuracy of the lifetime measurement of DIRAC. 
It is the aim of the present article to provide the necessary numerical 
setting.

{\bf 2.} The formation of the atom and its subsequent decay into two neutral 
pions is induced by isospin breaking effects in the underlying theory. In
the present framework (QCD including photons), these are the electromagnetic 
interactions and  the mass difference of the up and down quarks. In the 
following, it is useful to count $\alpha\simeq 1/137$ and $(m_d-m_u)^2$ as 
small parameters of order $\delta$. 

More than forty years ago, Deser {\it et al.}~\cite{Deser} derived the 
formula for 
the width of the $\pi^- p$ atom at leading order  in isospin symmetry 
breaking. Later in Refs.~\cite{Uretsky,Bilenky}, this result was adapted to  
the $\pi^+\pi^-$ atom\footnote{There are a few misprints in 
Eq.~(6) for the pionium decay rate in Ref.~\cite{Uretsky}. The correct 
result is displayed in Ref.~\cite{Bilenky}.}. 
In particular, it was shown that - again at leading order in isospin symmetry 
breaking effects - the width  $\Gamma_{2\pi^0}^{\rm LO}$ of pionium is 
proportional to the square of the difference $a_0-a_2$,
\eq\label{deser}
\Gamma_{2\pi^0}^{\rm LO}=\frac{2}{9}\,\alpha^3 p^\star (a_0-a_2)^2\,\,   ;
\,\, \,
p^\star=(M_{\pi^+}^2-M_{\pi^0}^2-
\frac{1}{4}M_{\pi^+}^2\alpha^2)^{1/2}.   
\en
At leading order in $\delta$, the momentum $p^\star$ becomes
$\sqrt{2M_{\pi^+}(M_{\pi^+}-M_{\pi^0})}$ - this is the expression
used in~\cite{Uretsky,Bilenky}. We prefer to use Eq.~(\ref{deser}),
because in this manner, one disentangles the kinematical corrections - 
due to the expansion of the square root - from  true dynamical ones. 

In our recent article~\cite{Bern}, we derived a general expression 
for the pionium lifetime, that is valid  at leading and 
next-to-leading order in isospin breaking,
\eq\label{our}
\Gamma_{2\pi^0}=\frac{2}{9}\,\alpha^3
p^\star{\cal A}^2(1+K)\per
\en
The quantities ${\cal A}$ and $K$ are expanded in powers of
$\delta$. In particular, it has been shown in~\cite{Bern} that\footnote{
We use throughout the Landau symbols $O(x)$ [$o(x)$] for quantities that
vanish like $x$ [faster than $x$] when $x$ tends to zero. Furthermore,
it is understood that this holds modulo logarithmic terms, i.e. we write also
$O(x)$ for $x\ln x$. 
}
\eq\label{eq21}
{\cal A}&=&-\frac{3}{32\pi}\,{\rm Re}A^{+-00}_{\rm thr}+o(\delta),
\en
where ${\rm Re}A^{+-00}_{\rm thr}$ is calculated as follows.
One evaluates  the relativistic scattering amplitude for the  
process $\pi^+\pi^-\rightarrow\pi^0\pi^0$ at order $\delta$ near 
threshold and removes the (divergent) Coulomb phase. The real part of this 
matrix element 
develops singularities that behave like $|{\bf p}|^{-1}$ and 
$\ln 2|{\bf p}|/M_{\pi^+}$ near threshold ($\bf{p}$ denotes the center of 
mass momentum of the charged pions). The 
remainder,  evaluated at the $\pi^+\pi^-$ threshold ${\bf p=0}$, equals
${\rm Re}A^{+-00}_{\rm thr}$. It contains terms of order $\delta^0$ and
$\delta$ and is normalized such that, in the isospin symmetry limit, 
${\cal A}=a_0-a_2$. Finally, the quantity $K$ starts at order 
$\alpha \ln\alpha$ - its explicit expression up to and including terms of
order $\delta$ is given by
\eq\label{K}
K&=&\frac{\Delta_\pi}{9M^2_{\pi^+}}\,(a_0+2a_2)^2
-\frac{2\alpha}{3}\,(\ln\alpha-1)\,(2a_0+a_2)+o(\delta)\co\nonumber\\[2mm]
\Delta_\pi&=& M^2_{\pi^+}-M^2_{\pi^0}\per
\en

In the derivation of Eqs.~(\ref{our}) - (\ref{K}), the
chiral expansion has not been used - details will be provided 
in a forthcoming publication~\cite{Big}. 

The following remarks are in order. First, it turns out that numerically, 
the correction factor $(1+K)$ is very close to unity, see below. The 
present uncertainties in the values of the scattering lengths
that enter the expression for $K$ are thus of no significance in this respect.
Second, it follows that the measurement of the pionium lifetime amounts to 
a precise measurement of the quantity ${\cal A}$, that is not affected by 
use of any chiral expansion.

{\bf 3.} In order  to extract $\pi\pi$ scattering lengths from
experimental information on the width, we invoke ChPT and
relate the amplitude ${\cal A}$ and the difference of the $S$-wave
scattering lengths $a_0-a_2$ 
order by order in the chiral expansion. This may be achieved 
as follows. First, we expand  ${\cal A}$  in powers of the isospin
breaking parameter $\delta$. Because the scattering amplitude
$\pi^+\pi^-\rightarrow \pi^0\pi^0$ in QCD does not contain terms linear 
in the quark mass difference $m_u-m_d$, we write
\eq\label{AK}
{\cal A}&=&a_0-a_2+h_1\,(m_d-m_u)^2+h_2\,\alpha + o(\delta),
\en
where $a_0$ and $a_2$ denote, {\it by definition}, the strong $\pi\pi$
scattering lengths evaluated in QCD at $e=0,~m_u=m_d$. The quark masses 
are tuned  such that the pion mass in the isospin-symmetric world coincides
with the charged pion mass. The values of other parameters are not changed
when performing this isospin symmetry limit. For illustration, the difference
$a_0-a_2$ at one loop is given by~\cite{ChPT}
\eq\label{a0a2}
a_0-a_2=\frac{9M^2_{\pi^+}}{32\pi F^2}\left(1+\frac{M^2_{\pi^+}}
{288\pi^2 F^2}\, \left\{33+8\bar l_1+16\bar l_2-3\bar l_3\right\}\right)
+O(M^6_{\pi^+})\co
\en
where $\bar{l}_i$ denotes the running coupling constant $l_i^r$ at scale
$\mu=M_{\pi^+}$, and $F$ stands for the pion decay constant in the chiral
limit, $F\sim 88~{MeV}$~\cite{ChPTlit}.  

Second, the coefficients $h_i$ are evaluated in the framework of chiral 
perturbation theory. This calculation is straightforward - one only needs to 
determine in the $\pi^+\pi^-\rightarrow \pi^0\pi^0$ amplitude
the terms proportional to $\alpha$ and $(m_d-m_u)^2$. Fortunately, the result
at order $e^2p^2$ is already available in the literature~\cite{Knecht}.
At this order, $h_1$ vanishes. Indeed, $h_1$ 
is at least of order $\hat{m}$ in any order of the chiral expansion, with 
$\hat{m}=(m_u+m_d)/2$. The expression for $h_2$ can be read off from 
amplitude given in Ref.~\cite{Knecht}. The result is
\eq\label{h2}
h_1&=&O(\hat{m})\nonumber\co\\[2mm]
h_2&=&h_\Delta+h_\gamma +O(\hat{m}^2)\co \nonumber\\[2mm]
h_\Delta&=&\frac{3\Delta_\pi^{e.m.}}{32\pi\alpha F^2}\,
\biggl\{1 + \frac{M_{\pi^+}^2}{12\pi^2 F^2}
\biggl [\frac{23}{8}+\bar l_1 + \frac{3}{4}\, \bar l_3 
\biggr ]\biggr\} \co\nonumber\\[2mm]  
h_\gamma&=&\frac{3 M_{\pi^+}^2}{256\pi^2 F^2}\,\, p(k_i) \per
\en
Here, $p(k_i)$ denotes a particular combination of electromagnetic low-energy
constants $k_i^r$ that occur in the effective Lagrangian of 
$SU(2)\times SU(2)$ at order $e^2p^2$~\cite{Knecht,Meissner}. Using the 
Lagrangian of Ref.~\cite{Knecht}, one has 
\eq\label{pki}  
\hspace{-.3cm}p(k_i)=-30 + 9\bar k_1 + 6\bar k_3 + 2\bar k_6 + \bar k_8
+\frac{4}{3}Z(\bar k_1+2\bar k_2+6\bar k_4+12\bar k_6-6\bar k_8)\co
\nonumber\\[2mm]
\en
and
\eq\label{Z}
\Delta^{e.m.}_\pi=\Delta_\pi|_{m_u=m_d}\, , \quad\quad 
Z=\frac{\Delta_\pi}{8\pi\alpha F^2}\biggr|_{m_u=m_d=0}\per
\en
The quantities $\bar{k}_i$ denote again the running couplings $k_i^r$
at scale $\mu=M_{\pi^+}$. Note that according to our counting, the quantity
$L_\pi=\ln(M_{\pi^+}^2/M_{\pi^0}^2)$ introduced in Ref.~\cite{Knecht}, is of
order $\delta$ and hence does not contribute to $h_2$.

{\bf 4.}
In the numerical analysis, it is convenient to relate the particular
combination of electromagnetic low-energy constants $\bar k_i$ that appears
in Eq.~(\ref{pki}), to the low-energy constants $K_i^r$ in the 
$SU(3)\times SU(3)$ version of ChPT. Estimates for the numerical 
values of $K_i^r$ are available in the 
literature~\cite{Baur,Moussallam,Bijnens}, whereas the $\bar k_i$ have not yet 
been determined to the best of our knowledge. The relation between 
$\bar k_i$ and
$K_i^r$ is straightforward to work out: one e.g. evaluates the amplitude
$\pi^+\pi^-\rightarrow\pi^0\pi^0$ and $M_{\pi^0}^2$ in the framework of 
$SU(3)\times SU(3)$ and expands the result at small momenta and small quark
masses, $p,m_u,m_d  \ll m_s$. In this limit, the expression for the 
$SU(3)\times SU(3)$ amplitude goes over into the one given in~\cite{Knecht}, 
provided that one sets 
\eq\label{mapping}
\hspace*{-0.2cm}
p(k_i)&=&P(K_i)-8Z\bar{l}_4 \co\nonumber\\[2mm]
\hspace*{-0.2cm}
P(K_i)&=&\frac{128\pi^2}{3}
\biggl(-6(K_1^r+K_3^r)+3K_4^r-5K_5^r+K_6^r+6(K_8^r+K_{10}^r+K_{11}^r)\biggr)
\nonumber\\[2mm]
&-&(18+28Z)\ln\frac{M^2_{\pi^+}}{\mu^2}-
2Z\left(\ln\frac{m_s B_0}{\mu^2} +1\right) -30\per
\en
Here, $B_0$ is related to the chiral condensate in 
$SU(3)\times SU(3)$ and $m_s$ denotes the strange quark mass~\cite{ChPTlit}. 
Furthermore, we have used that, at this
order in the low-energy expansion, $Z$ equals its $SU(3)\times SU(3)$ analog
$Z_0$. The couplings $K_i^r$ can be expressed~\cite{Moussallam} as a 
convolution of a QCD correlation function with the photon propagator, plus 
a contribution from QED counterterms. We have 
checked that $P(k_i)$ is independent of the QCD scale $\mu_0$ that must be
introduced in the QCD Lagrangian after taking into account electromagnetic
effects~\cite{Moussallam,Bijnens}\footnote{We thank B. Moussallam for
clarifying remarks concerning this point.}.

Using the relation between $F$ and $F_\pi$ in QCD~\cite{ChPTlit},
\eq\label{Fpi}
F_\pi=F\, \biggl(1+\frac{M_{\pi^+}^2}{16\pi^2F^2}\, \bar l_4
+O(\hat{m}^2)\biggr)\, ,
\en
the decay width of pionium can finally be expressed in the form
\eq\label{widthfin}
\Gamma_{2\pi^0}=\frac{2}{9}\, \alpha^3p^*\, (a_0-a_2+\epsilon)^2\, (1+K)\co
\en
where
\eq\label{epsfin}
\epsilon&=&\alpha h_\Delta + \alpha h_\gamma +\cdots\nonumber\\[2mm]
        &=&\frac{3\Delta_\pi^{e.m.}}{32\pi F_\pi^2}
\biggl\{1 + \frac{M_{\pi^+}^2}{12\pi^2 F_\pi^2}
\biggl [\frac{23}{8}+\bar l_1 + \frac{3}{4}\, \bar l_3+\frac{3}{2}\, \bar l_4 
\biggr ]\biggr\}
\nonumber\\[2mm]
&+&\frac{3 \alpha M_{\pi^+}^2}{256\pi^2 F_\pi^2}\, 
(P(K_i)-8Z\bar l_4)
\, +\cdots \per 
\en
The ellipses denote terms of $O(\hat{m}(m_d-m_u)^2,\alpha \hat{m}^2)$
and of $o(\delta)$. We expect these terms to give a negligible contribution to
$\epsilon$. Note that the term proportional to $\bar l_4$ cancels in
the sum $\alpha h_\Delta+\alpha h_\gamma$ at $O(e^2p^2)$.
The formulae (\ref{widthfin}) and (\ref{epsfin}) will be used below for the 
numerical analysis.

{\bf 5.} For the  numerical analysis of Eqs.~(\ref{widthfin}), (\ref{epsfin}), 
we  use the following values for the various quantities that occur in these
expressions. First, we recall that the non-electromagnetic part of the pion 
mass difference is tiny, of  order $\sim 0.1~{\rm MeV}$~\cite{Reports}. 
Therefore, we identify $\Delta_\pi^{e.m.}$  with the experimentally measured 
total shift $\Delta_\pi$. Similarly, the value of $Z$ is determined from 
Eq.~(\ref{Z}) using the observed value of $\Delta_\pi$. 
Further, in the calculations we replace $m_s B_0$ by
$M_{K^+}^2-M_{\pi^+}^2/2$, according to our definition of the isospin symmetry
limit. The values used for the low-energy constants are:
$F_\pi=93.2~{\rm MeV}$~\cite{ChPTlit,pipi-NUCL}, 
$\bar l_1=-2.3\pm 3.7$, $\bar l_2=6.0\pm 1.3$, 
$\bar l_3=2.9\pm 2.4$~\cite{ChPTlit}, 
$\bar l_4=4.4\pm 0.3$~\cite{Bijnens-scalar}. For $K_i^r(\mu)$, we use the
values given by Baur and Urech in  Ref.~\cite[Table 1]{Baur}: 
$K_1^r=-6.4,~
 K_3^r=6.4,~
 K_4^r=-6.2,~
 K_5^r=19.9,~
 K_6^r=8.6,~
 K_8^r=K_{10}^r=0,~
 K_{11}^r=0.6$ (in units of $10^{-3}$). 
We evaluate $P(K_i)$ at scale $\mu=M_\rho$.
Further, we attribute an uncertainty $2/16\pi^2$ - that stems from 
dimensional arguments - to each  $K_i^r$. The values of $K_i^r$
obtained both by Moussallam~\cite{Moussallam} and by Bijnens and 
Prades~\cite{Bijnens}, lie then within the uncertainties attributed. 
In the final expression for $\epsilon$, the uncertainties coming from 
$\bar l_i$ and $K_i^r$, are added quadratically. Finally, we use for $a_0$ and
$a_2$ the values corresponding to  set~2 in Ref.~\cite{pipi-NUCL}: 
$a_0=0.206$, $a_2=-0.0443$, without attributing any error. These scattering
lengths enter only the correction $K$ in (\ref{widthfin}), which is 
very small. 

Inserting all this in $\epsilon$, and omitting the terms of higher
order in the chiral expansion and in isospin violating effects
[indicated by the ellipses in (\ref{epsfin})], we arrive at
\eq\label{numerics}
\epsilon=(0.58\pm 0.16)\cdot 10^{-2}\, ,\quad\quad
K=1.07\cdot 10^{-2}\per
\en
The error  in this result includes {\it only} the uncertainty in the 
values of the low-energy constants $K_i^r$ and $\bar{l}_i$. Resonance 
saturation introduces a scale dependence in the final result for $\epsilon$. 
If saturation is assumed at $\mu=500$ MeV ($\mu=1$ GeV), we find 
$\epsilon=0.51\cdot 10^{-2}$ ($\epsilon=0.62\cdot10^{-2}$). 
Eqs.~(\ref{widthfin}) - (\ref{numerics}) are the main result of this letter.
We add the following remarks concerning this analysis.

i) The  term  $\alpha h_\Delta$ ($\alpha h_\gamma$) contributes with 
$0.51\cdot 10^{-2}$ ($0.07\cdot 10^{-2}$) to $\epsilon$, see 
Eq.~(\ref{epsfin}). Therefore, $\alpha h_\Delta$ is by far the dominant
effect in the improvement of the leading order formula (\ref{deser}).
The dominant contribution in $K$ stems from the logarithmic term 
$\sim \alpha\ln{\alpha}$.

ii) At one-loop order,  $\alpha h_\Delta$ can be written in the following form,
\eq\label{eps_Delta}
\alpha h_\Delta=\frac{\Delta^{e.m.}_\pi}{3M_{\pi^+}^2}\,
\biggl\{a_0-a_2+\frac{M_{\pi^+}^4}{256\pi^3 F_\pi^4}\,   
\biggl [9 + 4 (\bar l_1 - \bar l_2) + \frac{21}{4}\bar l_3\biggr] \biggr\}\per 
\en
For the values of the low-energy constants $\bar l_1$, $\bar l_2$ and 
$\bar l_3$ given above, the second term in the curly bracket
turns out to be very small, of order $-6\cdot 10^{-3}$. Consequently, the 
one-loop correction to the ratio $\alpha h_\Delta/(a_0-a_2)$ is tiny,
and the latter is dominated by the tree-level contribution 
$\Delta_\pi/3M_{\pi^+}^2$. 

iii) If one is willing to rely  on the numerical values of $a_0$ and $a_2$ 
displayed  above, one may evaluate the lifetime of pionium in
the ground state: $\tau=3.25\cdot10^{-15}~{s}$. The correction
to the lowest-order formula (\ref{deser}) by Deser {\it et al.} then becomes
$(\Gamma_{2\pi^0}-\Gamma^{\rm LO}_{2\pi^0})/\Gamma^{\rm LO}_{2\pi^0} = 0.058$.
We prefer not to attach any error to these numbers, because they are 
based on the numerical values of $a_0$ and $a_2$, and these are not yet known
with sufficient accuracy.

{\bf 6.} We now compare our result with recent work in the literature and 
start the discussion with  Refs.~\cite{Rasche}, where hadronic atoms have  
been studied in the framework of potential scattering theory. The correction
to the leading-order formula for the width has been worked out 
in~\cite{Rasche} numerically - a comparison with our analytic result is 
thus not possible. The numerical result quoted in~\cite{Rasche} differs 
significantly from ours: isospin violating corrections to the leading order 
result increase the lifetime according to these authors, in contrast to
Eqs.~(\ref{widthfin}), (\ref{numerics}). The discrepancy 
does not come as a surprise - in the present form, the 
potential approach~\cite{Rasche} does not reproduce all isospin-breaking 
terms that are present in the Standard Model. The leading
correction in ChPT stems from tuning the quark mass in the isospin-symmetric
phase such that the pion mass in the isospin-symmetric world coincides with the
charged pion mass. Since the pion-pion interaction depends on the quark mass,
this effectively leads to a change in the $\pi\pi$ potential
when the isospin limit is considered. In addition, the potential model
in~\cite{Rasche} does not take into account the direct quark-photon effects
encoded in the low-energy constants $\bar k_i$ in our approach. We
believe that, for a consistent calculation of the corrections to the
$\pi^+\pi^-$ atom decay width, the potential in the scattering theory approach
should be matched to ChPT in the {\it isospin violating}
phase. This procedure  would guarantee that the above mentioned
effects are included.

{\bf 7.}
In order to present a coherent comparison with other calculations performed
in the framework of Quantum Field Theory, it is useful to expand also the
quantity $K$, similarly to ${\cal A}$ in (\ref{AK}),
\eq
K=f_1\,(m_d-m_u)^2+f_2\,\alpha\ln\alpha+f_3\,\alpha + o(\delta),
\en
and to rewrite the decay width (\ref{our}) in the form
\eq\label{ourR}
\Gamma_{2\pi^0}&=&\frac{2}{9}\,\alpha^3\, p^\star R^2\, ,\nonumber\\[2mm]
R&=& a_0-a_2 +r_1\alpha\ln\alpha +r_2\alpha+r_3(m_d-m_u)^2+o(\delta)\, ,
\en
where the coefficients in the expansion are given by
\eq\label{r_i}
r_1&=&\frac{1}{2}\,(a_0-a_2) f_2\, ,\nonumber\\[2mm]
r_2&=&h_2+\frac{1}{2}\,(a_0-a_2) f_3\, ,\nonumber\\[2mm]
r_3&=&h_1+\frac{1}{2}\,(a_0-a_2) f_1\per
\en
The comparison with other approaches becomes then rather easy: one 
compares the analytic expressions for $r_i$ at a given order in the 
chiral expansion.

The correction to the pionium decay width were evaluated in 
Refs.~\cite{Sazdjian} by use of 3D constraint theory equations, 
and in Refs.~\cite{Atom} in the Bethe-Salpeter approach. 
The term proportional to $\alpha^2/\Delta_\pi$ that emerges
from the expansion of $p^\star$, was omitted in the final expressions
for the decay width in these papers. Note that though this term
is algebraically of order $\delta$, the numerical effect coming from it is
negligible. Further, rewriting the expressions found in these
works in the form (\ref{ourR},\ref{r_i}), we find that the
coefficient $r_1$ coincides to all orders in the chiral expansion with 
our result, whereas  $r_2$ coincides with the above result up to and 
including terms of order $p^2$. The matching relation (\ref{mapping})
does however not agree with the one proposed in Refs.~\cite{Sazdjian}.
The expressions in Refs.~\cite{Sazdjian} and~\cite{Atom} contain some of 
the higher-order terms in the quark mass expansion and in isospin breaking 
effects. These cannot be reliably predicted based on $O(e^2p^2)$ 
calculations alone. The effect of those terms is however small.
The numerical values of the corrections in 
\cite{Sazdjian,Atom} almost coincide with our result.

Labelle and Buckley~\cite{Labelle} were the first to apply the non 
relativistic effective Lagrangian approach to the problem of hadronic atoms.
They  evaluated  the leading order term, and in addition the 
correction due to electron vacuum polarization. 
This contribution is analytically of order $\alpha^2$ and thus beyond the 
accuracy considered here. Although this correction is potentially large,
because it is proportional to $(M_{\pi^+}/m_{electron})\alpha^2 \simeq
2\alpha$, it amounts to a tiny contribution $3.9\cdot 10^{-4}$
to $\epsilon$, which one may safely ignore. 
The vacuum polarization correction to the $\pi^+\pi^-$ atom decay width was
obtained independently in Ref.~\cite{Karshenboim} - the result agrees to that
of~\cite{Labelle}. Note that the part of 
the correction due to vacuum polarization, with no Coulomb corrections in the
intermediate state,  was considered also in 
Refs.~\cite{Atom}, and the result agrees with the corresponding one given in 
Ref.~\cite{Labelle}.

Recently, Kong and Ravndal~\cite{Kong,Kong_rel} and Holstein~\cite{Holstein} 
have applied non relativistic effective Lagrangian techniques to this problem
as well. In Refs.~\cite{Kong,Kong_rel}, Coulomb corrections are not taken into
account, which amounts to $r_1=0$. The relativistic correction found in 
Ref.~\cite{Kong_rel} amounts to rewriting the phase space factor 
$\sqrt{2M_{\pi^+}(M_{\pi^+}-M_{\pi^0})}$ as $\sqrt{M_{\pi^+}^2-M_{\pi^0}^2}$ 
- this term  is thus included in our approach. On the other hand, we do not 
agree in the value of the coefficient $r_2$ given in Ref.~\cite{Kong} 
already at leading order in the quark mass expansion. The discrepancy is 
due to the fact that - in~\cite{Kong} - the matching of the effective 
couplings in the non relativistic Lagrangian has not been carried out with 
a precision that is required to pin down all terms at this order in the 
expression for the width. If a complete matching at $O(\delta)$ is performed, 
agreement is achieved at the leading order in quark mass 
expansion~\cite{Kong-revised}. The result of Ref.~\cite{Holstein} 
for $r_2$  agrees with Ref.~\cite{Kong} - hence, it also shares its 
shortcomings.
The contribution corresponding to $r_1$ is omitted in the final expression 
for the width in~\cite[Eq.(95)]{Holstein}, although at previous stages in 
that work, the Coulomb corrections had been discussed. Finally, effects 
from $m_u \neq m_d$ are not disentangled in \cite{Kong,Kong_rel,Holstein}. 

In Ref.~\cite{Soto}, an effective non relativistic Lagrangian was used to 
derive the expression for the decay width in terms of effective couplings 
in that Lagrangian. The result agrees with ours~\cite[Eq.(11)]{Bern}.
The matching to the relativistic amplitude is carried out in~\cite{Soto} 
order by  order in the chiral expansion. We agree with the result 
of~\cite{Soto} for the coefficient $r_1$ at leading order in the chiral 
expansion. The result for the coefficient $r_2$ is not explicitly given 
in~\cite{Soto}, and a comparison is therefore not possible.

\vspace*{-0.1cm}

The result presented in Ref.~\cite{Bunatian} differs from ours in many respect.
In particular, the expression for the decay width given in 
Ref.~\cite{Bunatian} contains an ultraviolet divergence that is regularized by
introducing an explicit cutoff. A systematic renormalization procedure is
not discussed. For this reason, the result of~\cite{Bunatian} can not
be compared with the present one.  

\vspace*{-0.1cm}

Finally, we comment on the work of Ref.~\cite{Minkowski}.
First we note that in our approach, the decay width is calculated by 
considering the elastic $\pi^0\pi^0\rightarrow\pi^0\pi^0$ amplitude as a
function of the energy $E$ of the $\pi^0\pi^0$ pair. This amplitude
develops a pole at $E=E_p$ on the second Riemann sheet - the imaginary 
part Im($E_p$) is related to the width in the standard manner.
As far as we can see, the approaches proposed in~\cite{Kong,Kong_rel,Holstein} 
amount to the same definition of the width, because it is identified
in these works with the imaginary part of the level shift in 
Rayleigh-Schr\"odinger perturbation theory. On the other hand, 
Ref.~\cite{Minkowski} amounts to an alternative definition of 
$\Gamma_{2\pi^0}$ 
- a direct comparison with the present work  is therefore not possible.

{\bf 8.} In conclusion, we have analyzed analytically and numerically the 
general formula~\cite{Bern} for the decay width of the pionium ground state
at one-loop order in ChPT, and we have compared our result with other work 
available in the literature. We have in particular identified the reason 
for discrepancies of our result with other approaches. As we have shown, 
a precise determination of the pionium lifetime allows one to measure the 
amplitude ${\cal A}$ in Eq.~(\ref{our}), where no chiral expansion is used. 
By use of ChPT, one may determine the combination $a_0-a_2$ of $S$-wave 
scattering lengths from Eqs.~(\ref{widthfin}) - (\ref{numerics}), that 
constitute the main result of this letter. As we expect the corrections 
due to higher orders in the chiral expansion to be small, we infer from 
this result  that an accurate determination of $a_0-a_2$ from a precise 
lifetime measurement is indeed feasible.


{\it Acknowledgments}. We are grateful to M. Knecht, H. Leutwyler, 
L. L. Nemenov, F. Ravndal, H. Sazdjian, J. Schacher,  J. Soto,  
J. Stern and O. V. Tarasov for useful discussions. 
V.E.L. and A.R. acknowledge interesting discussions with the members of DIRAC
collaboration during their seminars at CERN.
This work was supported in part by the Swiss National 
Science Foundation, and by TMR, BBW-Contract No. 97.0131 
and EC-Contract No. ERBFMRX-CT980169 (EURODA$\Phi$NE).


\begin{thebibliography}{99}

\bibitem{Bern}A.~Gall, J.~Gasser, V.~E.~Lyubovitskij, and A.~Rusetsky, 
Phys. Lett. B 462 (1999) 335.

\bibitem{DIRAC}B.~Adeva {\it et al.}, CERN proposal CERN/SPSLC 95-1 (1995). 

\bibitem{Deser} S.~Deser, M.~L.~Goldberger, K.~Baumann, and 
W.~Thirring,  Phys. Rev. 96 (1954) 774.

\bibitem{Uretsky}J.~L.~Uretsky and T.~R.~Palfrey, Jr.,
Phys. Rev. 121 (1961) 1798. 

\bibitem{Bilenky}S.~M.~Bilenky, Van Kheu Nguyen, L.~L.~Nemenov, and 
F.~G.~Tkebuchava, Sov. J. Nucl. Phys. 10 (1969) 469. 

\bibitem{ChPTlit}J.~Gasser and H.~Leutwyler, Ann. Phys. (N.Y.) 158 (1984) 142;
Nucl. Phys. B 250 (1985) 465.

\bibitem{ChPT}J.~Gasser and H.~Leutwyler, Phys. Lett. B 125 (1983) 325;
J.~Bijnens, G.~Colangelo, G.~Ecker, J.~Gasser,  
and M.~E.~Sainio, Phys. Lett. B 374 (1996) 210.

\bibitem{Stern}M.~Knecht, B.~Moussallam, J.~Stern, and N.~H.~Fuchs, 
Nucl. Phys. B 457 (1995) 513; ibid. B 471 (1996) 445.

\bibitem{Big}A.~Gall, J.~Gasser, V.~E.~Lyubovitskij, and A.~Rusetsky,
in preparation.

\bibitem{Knecht}M.~Knecht and R.~Urech, Nucl. Phys. B 519 (1998) 329.  

\bibitem{Meissner} U.-G.~Mei\ss ner, G.~M\"uller, and S.~Steininger,
Phys. Lett. B 406 (1997) 154; ibid.  B 407 (1997) 454 (E).

\bibitem{Baur}R.~Baur and R.~Urech, Nucl. Phys. B 499 (1997) 319. 

\bibitem{Moussallam}B.~Moussallam, Nucl. Phys. B 504 (1997) 381. 

\bibitem{Bijnens}J.~Bijnens and J.~Prades, Nucl. Phys. B 490 (1997) 239.


\bibitem{Reports}J.~Gasser and H.~Leutwyler, Phys. Rep. 87 (1982) 77. 

\bibitem{pipi-NUCL}J.~Bijnens, G.~Colangelo, G.~Ecker, J.~Gasser, and
M.~E.~Sainio, Nucl. Phys. B 508 (1997) 263.

\bibitem{Bijnens-scalar}J.~Bijnens, G.~Colangelo, and P.~Talavera, 
JHEP 9805 (1998) 014.

\bibitem{Rasche}
U.~Moor, G.~Rasche, and W.~S.~Woolcock, 
Nucl. Phys. A 587 (1995) 747;
A.~Gashi, G.~C.~Oades, G.~Rasche, and W.~S.~Woolcock, 
Nucl. Phys. A 628 (1998) 101;
G.~Rasche and A.~Gashi, preprint hep-ph/9807564.  

\bibitem{Sazdjian}H.~Jallouli and H.~Sazdjian, Phys. Rev. D 58 (1998) 014011; 
H.~Sazdjian, preprint hep-ph/9809425. 

\bibitem{Atom} 
V.~E.~Lyubovitskij and A.~G.~Rusetsky, Phys. Lett. B  389 (1996) 181;
V.~E.~Lyubovitskij, E.~Z.~Lipartia, and A.~G.~Rusetsky, 
JETP Lett. 66 (1997) 783;
M.~A.~Ivanov, V.~E.~Lyubovitskij, E.~Z.~Lipartia, and A.~G.~Rusetsky, 
Phys. Rev. D 58 (1998) 094024. 

\bibitem{Labelle}P.~Labelle and K.~Buckley, preprint hep-ph/9804201. 

\bibitem{Karshenboim}U.~Jentschura, G.~Soff, V.~Ivanov, and 
S.~G.~Karshenboim, Phys. Lett. A 241 (1998) 351.   

\bibitem{Kong}X.~Kong and F.~Ravndal, Phys. Rev. D 59 (1999) 014031.

\bibitem{Kong_rel}X.~Kong and F.~Ravndal, preprint hep-ph/9905539.

\bibitem{Kong-revised}F.~Ravndal (private communication), and revised version 
of Ref.~\cite{Kong_rel}.

\bibitem{Holstein} 
B.~R.~Holstein, preprint nucl-th/9901041.

\bibitem{Soto}D.~Eiras and J.~Soto, preprint hep-ph/9905543. 

\bibitem{Bunatian}G.~G.~Bunatian, Nucl. Phys. A  645 (1999) 314. 

\bibitem{Minkowski} P.~Minkowski, preprint hep-ph/9808387. 



\end{thebibliography}
\end{document}